\documentclass[12pt]{article}
\usepackage{amsfonts}
\newcommand{\eqdef}{\stackrel{\rm def}{=}}

\begin{document}

\baselineskip=20pt

%%%%%%%%%%%%%%%%%%%%%%%%%%%%%%%%%%%%%%%%%%%%%%%%%%%%%%%%%%%%
%                                                          %
%  Title page                                              %
%                                                          %
%%%%%%%%%%%%%%%%%%%%%%%%%%%%%%%%%%%%%%%%%%%%%%%%%%%%%%%%%%%%
\newfont{\elevenmib}{cmmib10 scaled\magstep1}
\newcommand{\preprint}{
    \begin{flushleft}
      \elevenmib Yukawa\, Institute\, Kyoto\\
    \end{flushleft}\vspace{-1.3cm}
    \begin{flushright}\normalsize  \sf
      DPSU-04-2\\
      YITP-04-41\\
      {\tt hep-th/0407259} \\ July 2004
    \end{flushright}}
\newcommand{\Title}[1]{{\baselineskip=26pt
    \begin{center} \Large \bf #1 \\ \ \\ \end{center}}}
\newcommand{\Author}{\begin{center}
    \large \bf S.~Odake${}^a$ and R.~Sasaki${}^b$ \end{center}}
\newcommand{\Address}{\begin{center}
      $^a$ Department of Physics, Shinshu University,\\
      Matsumoto 390-8621, Japan\\
      ${}^b$ Yukawa Institute for Theoretical Physics,\\
      Kyoto University, Kyoto 606-8502, Japan
    \end{center}}
\newcommand{\Accepted}[1]{\begin{center}
    {\large \sf #1}\\ \vspace{1mm}{\small \sf Accepted for Publication}
    \end{center}}

\preprint
\thispagestyle{empty}
\bigskip\bigskip\bigskip

\Title{Polynomials Associated with Equilibria of\\
Affine Toda-Sutherland Systems}
\Author

\Address
\vspace{1cm}
\begin{abstract}
An affine Toda-Sutherland system is a {\em quasi-exactly solvable\/}
multi-particle dynamics based on an affine simple root system.
It is a `cross' between two well-known integrable multi-particle
dynamics, an affine Toda molecule and a Sutherland system.
Polynomials describing the equilibrium positions of affine
Toda-Sutherland systems are determined for all affine simple root
systems.

\end{abstract}

%\newpage
%%%%%%%%%%%%%%%%%%%%%%%%%%%%%%%%%%%%%%%%%%%%%%%%%%%%%%%%%%%%%%%
%                                                             %
%  1. Introduction                                            %
%                                                             %
%%%%%%%%%%%%%%%%%%%%%%%%%%%%%%%%%%%%%%%%%%%%%%%%%%%%%%%%%%%%%%%
\section{Introduction}
\label{intro}
\setcounter{equation}{0}

Given a multi-particle dynamical system, to find and describe
its equilibrium position has practical as well as theoretical
significance. As is well-known, near the equilibrium the system is
reduced to a collection of harmonic oscillators and that their
spectra give the exact order $\hbar$ part of the full quantum
spectra \cite{ls1}.
Naively, one could describe the equilibrium position by zeros of a
certain polynomial.
In this  way one obtains the celebrated classical orthogonal
polynomials for {\em exactly solvable\/} multi-particle dynamics.
For the Calogero systems \cite{Cal} based on the $A$ and $B$ ($C$,
$BC$ and $D$) root systems, the equilibrium positions correspond to the
zeros of the Hermite and Laguerre polynomials \cite{sti,szego,calmat,cs}.
For the Sutherland systems \cite{Sut} based on the $A$ and $B$ ($C$,
$BC$ and $D$) root systems, the equilibrium positions correspond to the
zeros of the Chebyshev and Jacobi polynomials \cite{cs}.
Polynomials describing the equilibria of the Calogero and Sutherland
systems based on the exceptional root systems are also determined
\cite{os1}.
In all these cases the frequencies of small oscillations at the
equilibrium are ``{\em quantised\/}" \cite{cs, ls2}.
For another family of multi-particle dynamics based on root
systems, the Ruijsenaars-Schneider systems \cite{Ruij-Sch},
which are {\em deformation\/} of the Calogero and Sutherland
systems, the corresponding polynomials are determined \cite{rs,os2}.
They turn out to be {\em deformation\/} of the Hermite, Laguerre
and Jacobi polynomials which inherit the orthogonality \cite{os2}.
The frequencies of small oscillations at the
equilibrium are also ``{\em quantised\/}" \cite{rs}.
Another interesting feature is that the equations determining the
equilibrium look like {\em Bethe ansatz\/} equations.

One is naturally led to a similar investigation for partially
solvable or {\em quasi-exactly solvable} \cite{turb}
multi-particle dynamics.
{}From a not-so-long list of known quasi-exactly solvable
multi-particle dynamical systems \cite{st1}, we pick up the
so-called affine Toda-Sutherland systems \cite{kls} and determine
polynomials  describing the equilibrium positions.
These polynomials, as well as all the polynomials mentioned above,
are characterised as having {\em integer\/} coefficients only.

%%%%%%%%%%%%%%%%%%%%%%%%%%%%%%%%%%%%%%%%%%%%%%%%%%%%%%%%%%%%%%%
%                                                             %
%  2. affine-Toda-Sutherland systems                          %
%                                                             %
%%%%%%%%%%%%%%%%%%%%%%%%%%%%%%%%%%%%%%%%%%%%%%%%%%%%%%%%%%%%%%%
\section{affine Toda-Sutherland systems}

The affine Toda-Sutherland systems are quasi-exactly solvable
\cite{turb}  multi-particle dynamics based on any  crystallographic
root system. Roughly speaking, they are obtained by `crossing' two
well-known integrable dynamics, the affine-Toda molecule and the
Sutherland system.
Given a set of {\em affine simple roots\/}
$\Pi_0=\{\alpha_0,\alpha_1,\ldots,\alpha_r\}$,
$\alpha_j\in{\mathbb{R}}^r$, let us introduce a {\em prepotential\/}
$W$ \cite{bms}
\begin{equation}
   W(q)=g\sum_{j=0}^rn_j\log|\sin(\alpha_j\cdot q)|,
   \quad q={}^t(q_1,\ldots,q_r)\in{\mathbb{R}}^r,
   \label{prep}
\end{equation}
in which $g$ is a positive coupling constant and $\{n_j\}$ are the
Dynkin-Kac labels for $\Pi_0$.
That is, they are the integer coefficients of the affine simple root
$\alpha_0$; $-\alpha_0=\sum_{j=1}^r n_j\alpha_j$, $n_0\equiv 1$.
For simply-laced and un-twisted non-simply laced affine root systems
$\alpha_0$ is the lowest long root, whereas for twisted non-simply laced
affine root systems, $\alpha_0$ is the lowest short root. In either case
$h\eqdef \sum_{j=0}^r n_j$ is the {\em Coxeter number\/}.
This leads to the classical Hamiltonian
\begin{equation}
   H_C={1\over2}\sum_{j=1}^rp_j^2+{1\over2}\sum_{j=1}^r\left({\partial
   W(q)\over{\partial q_j}}\right)^2.
   \label{casham}
\end{equation}
It is shown \cite{kls} that the equilibrium position $\bar{q}$ is given
by a {\em universal\/} formula in terms of the dual Weyl vector
$\varrho^\vee$:
\begin{equation}
   {\partial W(\bar{q})\over{\partial q_j}}=0\quad
   \Leftrightarrow \quad
   \bar{q}={\pi\over h}\varrho^\vee,\qquad
   \varrho^\vee\eqdef\sum_{j=1}^r\lambda_j^\vee.
\end{equation}
The dual fundamental weight $\lambda_j^\vee$ is defined in terms of the
fundamental weight $\lambda_j$ by
$\lambda_j^\vee\eqdef ({2/{\alpha_j^2}})\lambda_j$, which satisfies
$\alpha_j\cdot\lambda^\vee_k=\delta_{j\,k}$.
At the equilibrium, the classical multi-particle dynamical system
(\ref{casham}) is reduced to a set of harmonic oscillators.
The frequencies (not frequencies squared) of small oscillations at the
equilibrium of the affine Toda-Sutherland model are given up to the
coupling constant $g$ by \cite{kls}
\[
   {1\over{\sin^2{\pi\over h}}}\left\{m_1^2,m_2^2,\ldots,m_r^2\right\},
\]
in which $m_j^2$ are the so-called affine Toda masses \cite{bcds}.
Namely, they are the eigenvalues of a symmetric $r\times r$ matrix $M$,
\(
   M_{kl}=\sum_{j=0}^rn_j(\alpha_j)_k(\alpha_j)_l
\), or
\(
   M=\sum_{j=0}^rn_j\alpha_j\otimes\alpha_j
\),
which encode the integrability of affine Toda field theory.
In \cite{bcds} it is shown for the non-twisted cases that
the vector $\mathbf{m}={}^t(m_1,\ldots,m_r)$, if ordered properly,
is the {\em Perron-Frobenius\/} eigenvector of the incidence matrix
(the Cartan matrix) of the corresponding root system.

\bigskip
The corresponding {\em quantum\/} Hamiltonian \cite{ls1,bms} is
\begin{equation}
   H_Q={1\over2}\sum_{j=1}^rp_j^2+{1\over2}\sum_{j=1}^r\left[\left({\partial
   W(q)\over{\partial q_j}}\right)^2+{\partial^2
   W(q)\over{\partial q_j^2}}\right],
   \label{quham}
\end{equation}
which is partially solvable or {\em quasi-exactly solvable\/}
for some affine simple root systems.
Namely for $A_{r-1}^{(1)}$, $D_3^{(1)}$, $D_{r+1}^{(2)}$, $C_{r}^{(1)}$
and $A_{2r}^{(2)}$, the above Hamiltonian (\ref{quham}) is known to have
a few exact eigenvalues and corresponding exact eigenfunctions \cite{kls}.

The polynomials related to the equilibrium position $\bar{q}$ are
easy to define for the classical root systems, $A$, $B$, $C$ and $D$.
As in the Sutherland cases, we introduce a polynomial having zeros at
$\{\sin \bar{q}_j\}$ or $\{\cos2\bar{q}_j\}$:
\begin{equation}
   P_r(q)\propto \prod_{j=1}^r(x-\sin\bar{q}_j),\quad
   \prod_{j=1}^r(x-\cos2\bar{q}_j).
   \label{polyform1}
\end{equation}
For the exceptional root systems, let us choose a set of $D$ vectors
${\cal R}$
\[
   {\cal R}=\{\mu^{(1)},\ldots,\mu^{(D)}\,|\,\mu^{(a)}\in {\mathbb R}^r\},
\]
which form a single orbit of the corresponding Weyl group.
For example, they are the set of roots $\Delta$ itself for simply laced
root systems, the set of long (short, middle) roots $\Delta_L$ ($\Delta_S$,
$\Delta_M$) for non-simply laced root systems and the so-called sets of
{\em minimal weights\/}.
The latter is better specified by the corresponding fundamental
representations, which are all the fundamental representations of $A_r$,
the vector ($\bf V$), spinor ($\bf S$) and conjugate spinor ($\bar{\bf S}$)
representations of $D_r$ and $\bf 27$ ($\overline{\bf 27}$) of $E_6$
and $\bf 56$ of $E_7$.
By generalising the above examples (\ref{polyform1}), we define
polynomials
\begin{equation}
   P_{\Delta}^{\cal R}(x)\propto
   \prod_{\mu\in{\cal R}}\Bigl(x-\sin(\mu\cdot\bar{q})\Bigr),\quad
   \prod_{\mu\in{\cal R}}\Bigl(x-\cos(2\mu\cdot\bar{q})\Bigr).
  \label{polyform2}
\end{equation}
For more general treatment we refer to our previous article \cite{os1}.

\bigskip
The resulting polynomials for various affine root systems $\Pi_0$ are
(we follow the affine Lie algebra notation used in \cite{kls,bcds}):
%%%%%%%%%%%
%  A      %
%%%%%%%%%%%
\paragraph{$A_{r-1}^{(1)}$ :}
In this case the equilibrium position is exactly the same as that of the
$A_{r-1}$ Sutherland \cite{Sut} and $A_{r-1}$ Ruijsenaars-Sutherland
system \cite{os2},
\(
   \bar{q}=({\pi/2h})\,{}^t(r-1,r-3,\ldots,-(r-1))
\)
with $h=r$.
Thus the polynomial is also the same, the Chebyshev polynomial of the
first kind:
\(
   2^{r-1}\prod_{j=1}^{r}\Bigl(x-\sin\bar{q}_j\Bigr)
   =T_{r}(x)=\cos r\varphi
\),
\(x=\cos\varphi\).

%%%%%%%%%%%
%  B,D,A  %
%%%%%%%%%%%
\paragraph{$B_r^{(1)}$ \& $D_{r+1}^{(2)}$  \& $A_{2r}^{(2)}$ :}
The Coxeter number is $h=2r$ for $B_r^{(1)}$, $h=r+1$ for
$D_{r+1}^{(2)}$ and $h=2r+1$ for  $A_{2r}^{(2)}$.
The equilibrium position is equally spaced
\(
   \bar{q}=({\pi/h})\,{}^t(r,r-1,\ldots,1).
\)
We obtain the Chebyshev polynomial of the second kind,
\(
   U_n(x)={\sin(n+1)\varphi/{\sin\varphi}}
\),
\(
   x=\cos\varphi,
\) for $B_r^{(1)}$ and a product of them for $D_{r+1}^{(2)}$ and
a sum of them for $A_{2r}^{(2)}$,
\begin{equation}
   2^{r-1}\prod_{j=1}^r(x-\cos2\bar{q}_j)=
   \left\{
   \begin{array}{ll}
    (x+1)U_{r-1}(x),\quad & B_r^{(1)},\\[6pt]
    (x+1)U_{r/2}(x)U_{(r-2)/2}(x)+1/2,
    &D_{r+1}^{(2)},\quad r:\ \mbox{even},\\[6pt]
    (x+1)U_{(r-1)/2}(x)^2,
    &D_{r+1}^{(2)},\quad r:\ \mbox{odd},\\[6pt]
    \left(U_{r}(x)+U_{r-1}(x)\right)/2, &A_{2r}^{(2)}.
   \end{array}
   \right.
\label{bpol}
\end{equation}

%%%%%%%%%%%
%  C,A    %
%%%%%%%%%%%
\paragraph{$C_r^{(1)}$ \& $A_{2r-1}^{(2)}$ :}
The Coxeter number is $h=2r$ for $C_r^{(1)}$ and $h=2r-1$ for
$A_{2r-1}^{(2)}$. The equilibrium position is equally spaced
\(
   \bar{q}=({\pi/2h})\,{}^t(2r-1,2r-3,\ldots,3,1).
\)
We obtain the Chebyshev polynomial of the first kind $T_r(x)$ for
$C_r^{(1)}$ and a sum of them for $A_{2r-1}^{(2)}$,
\begin{equation}
   2^{r-1}\prod_{j=1}^r(x-\cos2\bar{q}_j)=
   \left\{
   \begin{array}{ll}
     \, T_{r}(x),\quad & C_r^{(1)},\\[6pt]
     T_{r}(x)+T_{r-1}(x),&A_{2r-1}^{(2)}.
   \end{array}
   \right.
\label{cpol}
\end{equation}

%%%%%%%%%%%
%  D      %
%%%%%%%%%%%
\paragraph{$D_r^{(1)}$ :}
The Coxeter number is $h=2(r-1)$ and the equilibrium position
is equally spaced
\(
   \bar{q}=({\pi/ h})\,{}^t(r-1,r-2,\ldots,1,0).
\)
We obtain the Chebyshev polynomial of the second kind
\begin{equation}
   2^{r-2}\prod_{j=1}^r(x-\cos2\bar{q}_j)=(x^2-1) U_{r-2}(x).
\label{dpol}
\end{equation}

%%%%%%%%%%%
%  E_6    %
%%%%%%%%%%%
\paragraph{$E_6^{(1)}$ :}
The Coxeter number is $h=12$ and the equilibrium position is not equally
spaced
\(
   \bar{q}=({\pi/h})\,{}^t(4\sqrt{3},4,3,2,1,0).
\)
We consider the set of minimal weights $\mathbf{27}$ and the set of
positive roots $\Delta_+$, which consists of 36 roots.
The polynomials are
\begin{eqnarray}
   2^{20}\prod_{\mu\in\mathbf{27}}\Bigl(x-\sin(\mu\cdot\bar{q})\Bigr)
   &\!\!=\!\!&
   (-1+x)\,x^3\,(1+x)\,(-1+2x)^2\,(1+2x)^2\,(-1+2x^2)^2\nonumber\\[-8pt]
   &&\times(-3+4x^2)^3\,(1-16x^2+16x^4)^2,
\label{e6minpol}\\[6pt]
   2^{27}\prod_{\mu\in\Delta_+}\Bigl(x-\cos(2\mu\cdot\bar{q})\Bigr)
   &\!\!=\!\!&
   x^6\,(1+x)^3\,(-1+2x)^6\,(1+2x)^7\,(-3+4x^2)^7.
\label{e6root}
\end{eqnarray}

%%%%%%%%%%%
%  E_7    %
%%%%%%%%%%%
\paragraph{$E_7^{(1)}$ :}
The Coxeter number is $h=18$ and the equilibrium position is not equally
spaced
\(
   \bar{q}=({\pi/2h})\,{}^t(17\sqrt{2},10,8,6,4,2,0).
\)
We consider the set of minimal weights $\mathbf{56}$ and the set of
positive roots $\Delta_+$, which consists of 63 roots. The $\mathbf{56}$
is even, {\em ie\/} if $\mu\in\mathbf{56}$ then $-\mu\in\mathbf{56}$.
The positive part of $\mathbf{56}$ is denoted as $\mathbf{56}_+$.
The polynomials are
\begin{eqnarray}
   2^{24}\prod_{\mu\in\mathbf{56}_+}\Bigl(x-\cos(2\mu\cdot\bar{q})\Bigr)
   &\!\!=\!\!&
   x^4\,(-3 +4x^2)^3\,(-3+36x^2-96x^4+64x^6)^3,\\
   2^{59}\prod_{\mu\in\Delta_+}\Bigl(x-\cos(2\mu\cdot\bar{q})\Bigr)
   &\!\!=\!\!&
   (1+x)^4\,(-1+2x)^7\,(1+2x)^7\nonumber\\[-6pt]
   &&\times (-1+6x+8x^3)^8\,(1-6x+8x^3)^7.
\end{eqnarray}

%%%%%%%%%%%
%  E_8    %
%%%%%%%%%%%
\paragraph{$E_8^{(1)}$ :}
The Coxeter number is $h=30$ and the equilibrium position is not equally
spaced
\(
   \bar{q}=({\pi/h})\,{}^t(23,6,5,4,3,2,1,0).
\)
We consider  the set of positive roots $\Delta_+$, which consists of 120
roots. The polynomial is
\begin{eqnarray}
   &&2^{116}\prod_{\mu\in\Delta_+}\Bigl(x-\cos(2\mu\cdot\bar{q})\Bigr)
   =\nonumber\\
   &&\qquad\qquad (1+x)^4\,(-1+2x)^8\,(1+2x)^8\,
   (-1-2x+4x^2)^8\,(-1+2x+4x^2)^8\nonumber\\
   &&\qquad\qquad \times
   (1+8x-16x^2-8x^3+16x^4)^8\,(1-8x-16x^2+8x^3+16x^4)^9.
\end{eqnarray}

%%%%%%%%%%%
% F_4,E_6 %
%%%%%%%%%%%
\paragraph{$F_4^{(1)}$ \& $E_{6}^{(2)}$ :}
The Coxeter number is $h=12$ for $F_4^{(1)}$ and $h=9$ for $E_{6}^{(2)}$
and the equilibrium position is not equally spaced
\(
   \bar{q}=({\pi/h})\,{}^t(8,3,2,1).
\)
We consider the set of long positive roots $\Delta_{L+}$ and short
positive roots $\Delta_{S+}$, both of which consist of 12 roots
reflecting the self-duality of $F_4$ Dynkin diagram.
The polynomials for $F_4^{(1)}$ are
\begin{eqnarray}
   2^{9}\prod_{\mu\in\Delta_{S+}}\Bigl(x-\cos(2\mu\cdot\bar{q})\Bigr)
   &\!\!=\!\!&
   x^2(1+x)\,(-1+2x)^2\,(1+2x)^3\,(-3+4x^2)^2,
\label{f4spol}\\
   2^{9}\prod_{\mu\in\Delta_{L+}}\Bigl(x-\cos(2\mu\cdot\bar{q})\Bigr)
   &\!\!=\!\!&
   x^2(1+x)\,(-1+2x)^2\,(1+2x)\,(-3+4x^2)^3.
\label{f4lpol}
\end{eqnarray}
The polynomials associated with the twisted affine root system
$E_{6}^{(2)}$ are
\begin{eqnarray}
   2^{12}\prod_{\mu\in\Delta_{S+}}\Bigl(x-\cos(2\mu\cdot\bar{q})\Bigr)
   &\!\!=\!\!&
   (1+2x)^3\,(1-6x+8x^3)^3,\\
   2^{12}\prod_{\mu\in\Delta_{L+}}\Bigl(x-\cos(2\mu\cdot\bar{q})\Bigr)
   &\!\!=\!\!&
   2\,(-1+x)\,(1+2x)^2\,(1-6x+8x^3)^3.
\end{eqnarray}

%%%%%%%%%%%
% G_2,D_4 %
%%%%%%%%%%%
\paragraph{$G_2^{(1)}$ \& $D_{4}^{(3)}$ :}
The Coxeter number is $h=6$ for $G_2^{(1)}$ and $h=4$ for $D_{4}^{(3)}$
and the equilibrium position is
\(
   \bar{q}=({\pi/2h})\,{}^t(3\sqrt{6},\sqrt{2}).
\)
We consider the set of long positive roots $\Delta_{L+}$ and short
positive roots $\Delta_{S+}$, both of which consists of 3 roots
reflecting the self-duality of $G_2$ Dynkin diagram.
The polynomials for the untwisted $G_2^{(1)}$ are
\begin{eqnarray}
   2^3\prod_{\mu\in\Delta_{S+}}\Bigl(x-\cos(2\mu\cdot\bar{q})\Bigr)
   &\!\!=\!\!&
   2\,(1+x)\,(-1+2x)\,(1+2x),
\label{g2spol}\\
   2^3\prod_{\mu\in\Delta_{L+}}\Bigl(x-\cos(2\mu\cdot\bar{q})\Bigr)
   &\!\!=\!\!&
   (-1+2x)^2\,(1+2x).
\end{eqnarray}
The polynomials for the twisted $D_{4}^{(3)}$ are
\begin{eqnarray}
   \prod_{\mu\in\Delta_{S+}}\Bigl(x-\cos(2\mu\cdot\bar{q})\Bigr)
   &\!\!=\!\!&
   x^2\,(1+x),\\
   \prod_{\mu\in\Delta_{L+}}\Bigl(x-\cos(2\mu\cdot\bar{q})\Bigr)
   &\!\!=\!\!&
   x^2\,(-1+x).
\end{eqnarray}

%%%%%%%%%%%%%%
%  foldings  %
%%%%%%%%%%%%%%
\bigskip
Before closing this paper, let us briefly remark on the identities
arising from {\em foldings\/} of root systems.
Among them those relating two un-twisted root systems,
{\em ie\/} with superscript $(1)$ are quite simple.

\paragraph{Folding $A_{2r-1}^{(1)}\to C_r^{(1)}$ :}
The vector weights of $A_{2r-1}$ ($2r$ dim.) become those of $C_r$
($2r$ dim.).
This relates $T_{2r}$ to $T_{r}$ in (\ref{cpol}) as
\begin{equation}
  A_{2r-1}:\quad T_{2r}(x)=(-1)^r\, T_r(1-2x^2),\quad C_r^{(1)}.
\end{equation}

\paragraph{Folding $D_{r+1}^{(1)}\to B_r^{(1)}$ :}
This gives a quite obvious relation as seen from (\ref{dpol}) and (\ref{bpol}).

\paragraph{Folding $E_{6}^{(1)}\to F_4^{(1)}$ :}
In this folding the minimal weights $\mathbf{27}$ of $E_6$ become 
$\Delta_S$ (24 dim.) of $F_4$ plus three zero weights. Thus we obtain
\begin{equation}
  E_6^{(1)}:\quad 2\,(\ref{e6minpol})/x^3=(\ref{f4spol})_{x\to 1-2x^2}, \quad 
  F_4^{(1)}.
\end{equation}
We also obtain
\begin{equation}
  E_6^{(1)}:\qquad (\ref{e6root})=(\ref{f4spol})^2\times (\ref{f4lpol}), \qquad
  F_4^{(1)},
\end{equation}
since the 72 roots of $E_6$ are decomposed into $2\Delta_S+\Delta_L$ 
(24 dim.) of $F_4$.

\paragraph{Folding $D_{4}^{(1)}\to G_2^{(1)}$ :}
The vector weights of $D_4$ (8 dim.) decompose into  $\Delta_S$ (6 dim.) 
plus two zero weights of $G_2$ leading to the identity
\begin{equation}
  D_4^{(1)}:\qquad 2\,(\ref{dpol})_{r=4}/(x-1)=(\ref{g2spol}), \quad G_2^{(1)}.
\end{equation}

%%%%%%%%%%%%%%%%%%%%%%%%%%%%%%%%%%%%%%%%%%%%%%%%%%%%%%%%%%%%%%%
%                                                             %
%  Acknowledgments                                            %
%                                                             %
%%%%%%%%%%%%%%%%%%%%%%%%%%%%%%%%%%%%%%%%%%%%%%%%%%%%%%%%%%%%%%%
\section*{Acknowledgements}
S. O. and R. S. are supported in part by Grant-in-Aid for Scientific
Research from the Ministry of Education, Culture, Sports, Science and
Technology, No.13135205 and No. 14540259, respectively.

%%%%%%%%%%%%%%%%%%%%%%%%%%%%%%%%%%%%%%%%%%%%%%%%%%%%%%%%%%%%%%%%
%                                                             %
%  References                                                 %
%                                                             %
%%%%%%%%%%%%%%%%%%%%%%%%%%%%%%%%%%%%%%%%%%%%%%%%%%%%%%%%%%%%%%%

\end{document}